\documentclass[prb,reprint,showpacs,floatfix,epsf,epsfig,twocolumn]{revtex4} 
\usepackage{epsf}
\usepackage{epsfig}
\usepackage{epstopdf}
\usepackage{graphicx}

\begin{document}
\title{Ab initio Study of Effect of Co Substitution on the Magnetic Properties of Ni and Pt-based Heusler Alloys}
\author{Tufan Roy$^{1,2}$\footnote{Electronic mail: tufanroyburdwan@gmail.com}, Aparna Chakrabarti$^{1,2}$}
\affiliation{$^{1}$ Theory and Simulations Lab, HRDS, Raja Ramanna Centre for Advanced
Technology,  Indore-452013, India}
\affiliation{$^{2}$ Homi Bhabha National Institute, Training School Complex, Anushaktinagar, Mumbai-400094, India}

\begin{abstract} 
Using  density functional theory based calculations, we have carried out in-depth studies of effect of Co substitution on the magnetic properties of  Ni and Pt-based shape memory alloys.  We show the systematic variation of the total magnetic moment, as a function of Co doping.  A detailed analysis of evolution of Heisenberg exchange coupling parameters as a function of Co doping has been presented here. The strength of RKKY type of exchange interaction is found to decay with the increase of Co doping.
\end{abstract}


\maketitle

\section{Introduction}  

Full Heusler alloys (with typical formula A$_{2}$BC) have drawn considerable attention of the researchers over the last decades because of their possible technological applications. Upon cooling, some of the Heusler alloys undergo a structural transition from a high temperature cubic phase, namely austenite phase to a lower symmetry phase, called martensite phase below a certain temperature. This type of structural transition is referred as martensite transition, and the particular temperature at which the transition takes place is called martensite transition temperature. Ni$_{2}$MnGa belongs to this category of Heusler alloys.\cite{phil-web-1984,apl-ullakko-1996,apl-sozinov-2002} The Heusler alloys of this category may find their application as various devices, such as actuators, antenna, sensors etc. For the application purpose it is always desired that martensitic transition temperature is above the room temperature. In case of conventional shape memory effect, which is governed by the temperature, the actuation process is much more slow compared to a magnetically controlled actuation. So it is desirable to have a magnetic shape memory alloy with the Curie temperature (T$_{C}$) higher than the room temperature. It has been observed that both  T$_{M}$ and T$_{C}$ values are very much dependent on the composition of a particular Heusler alloy.\cite{prb-sroy-2009,prb-kataoka-2010, prb-achakrabarti-2005,prb-barman-2008,apl-achakrabarti-2009,jpcm-khan-2004,apl-mario-2011,apl-stadler-2006,prb-achakrabarti-2013,jalcom-troy-2015,prb-troy-2016,jmmm-troy-2016,jmmm-troy-2017}

There is also another category of full Heusler alloys, which are known to be metallic for one kind of spin channel and insulator for the other kind of spin channel because of their very high spin polarization (HSP) at the Fermi level. They are often called as half metallic Heusler alloys.\cite{prl-Groot-1983} Most of the Co-based Heusler alloys, like Co$_{2}$MnSn, Co$_{2}$MnGa belong to this category.\cite{JPD40HCK,PRB-76-024414-2007} These Heusler alloys may have potential application in spintronic devices.

Apart from the technological application, these Heusler alloys are very interesting because of their wide diversity in terms of magnetic property.
These alloys may be ferromagnetic, ferrimagnetic, anti-ferromagnetic and also non-magnetic depending on the chemical composition. So it is of immense interest to have an in depth study on the magnetic interactions present in these systems. In most of the full Heusler alloys, A$_{2}$BC, $B$ is the primary moment carrying atom, in many of the Heusler alloys, A$_{2}$BC, there is presence of a delocalized-like common d-band formed by the d-electrons of the $A$ and $B$ atoms, which are both typically first-row transition metal atoms.\cite{PRB28JK}  Additionally, there is also an indirect RKKY-type exchange mechanism\cite{RKKY} between the $B$ atoms, primarily mediated by the electrons of the $C$ atoms, which also plays an important role in defining the magnetic properties of these materials.\cite{PRB28JK,prb-sasioglu-2008}
Staunton et al\cite{staunton-jpcssp-1988} reported the role of RKKY interaction behind the origin of magnetic anisotropy of a system. For the magnetic shape memory alloys, magnetic anisotropy energy plays an important role.  In this regard also, it will be interesting to study of RKKY interaction in detail in these systems. 

In a very recent paper, we have shown in detail the similarities and differences between the Heusler alloys which are likely to show shape memory alloy (SMA) property and  which are not, in terms of the electronic, magnetic as well as mechanical properties.\cite{prb-troy-2016} In this paper, we focus our interest as to how the magnetic exchange interactions, mainly the RKKY type of interaction between the $B$ atoms of the A$_{2}$BC systems, are evolving in going from the materials which are prone to martensite transition (which are generally metallic in nature) to the other class of Heusler alloys (which are typically half-metallic in nature) i.e. which do not show SMA property. Here we study about the nature of RKKY types of interaction for four sets of materials Ni$_{2-x}$Co$_{x}$MnGa, Ni$_{2-x}$Co$_{x}$FeGa, Pt$_{2-x}$Co$_{x}$MnGa, Pt$_{2-x}$Co$_{x}$MnSn as a function of x (x=0.00, 0.25, 0.50, 0.75, 1.25, 1.50, 1.75, 2.00). In all the cases the material is likely to show SMA property for x=0.00 and is predicted to be half-metallic for x=2.00.
In the section following the methodology, the results of the work and the relevant discussion are presented. Finally, we summarize and conclude in the last section.
\section{Method}  

The Heusler alloys (A$_{2}$BC) studied here possess L2$_{1}$ structure that consists of four interpenetrating face-centered-cubic (fcc) sub-lattices with origin at fractional positions, (0.25, 0.25, 0.25), (0.75, 0.75, 0.75), (0.5, 0.5, 0.5), and (0.0, 0.0, 0.0). For the conventional Heusler alloy structure, the first two sub-lattices are occupied by $A$ atom and the third by $B$ and fourth by $C$ atom. In total, there are 16 atoms in the cell. While we study the Co substitution in A$_{2}$BC systems, the Co atom substitutes the $A$ atom only. First we carry out full geometry optimization of the materials, of all the materials corresponding to x=0.00, 0.25, 1.75, 2.00, using the 16 atom cell. For the geometry optimization, we employ the Vienna Ab Initio Simulation Package (VASP)\cite{prb-kreese-1996} in combination with the projector augmented wave  method.\cite{prb-blochl-1994} We use an energy cut-off of minimum 500 eV for the planewave basis set. The calculations have been performed with a  $k$ mesh of 15$\times$15$\times$15. The energy and force tolerance used were 10 $\mu$eV and 10 meV/\AA, respectively. After obtaining the equilibrium lattice constants of the four above-mentioned materials by using the VASP package we plot the same. A linear variation of the lattice constant  is observed.
We deduce the lattice constants of the other materials, corresponding to x=0.50, 0.75, 1.25, 1.50 by the method of interpolation. To gain insight into the magnetic interactions of these materials, we calculate and discuss their Heisenberg exchange coupling parameters. We use the Spin-polarized-relativistic Korringa-Kohn-Rostoker method (SPR-KKR) to calculate the Heisenberg exchange coupling parameters, Jij as implemented in the SPR-KKR programme package.\cite{rep-ebert-2011} The mesh of $k$ points for the SCF cycles has been taken as 21$\times$21$\times$21 in the BZ. The angular momentum expansion for each atom is taken such that lmax=3. The partial and total moments have also been calculated for all the materials studied. We use local density approximation (LDA) for exchange correlation functional.\cite{LDA}
\begin{figure}[ht*]
\includegraphics[width=8cm, height=8cm]{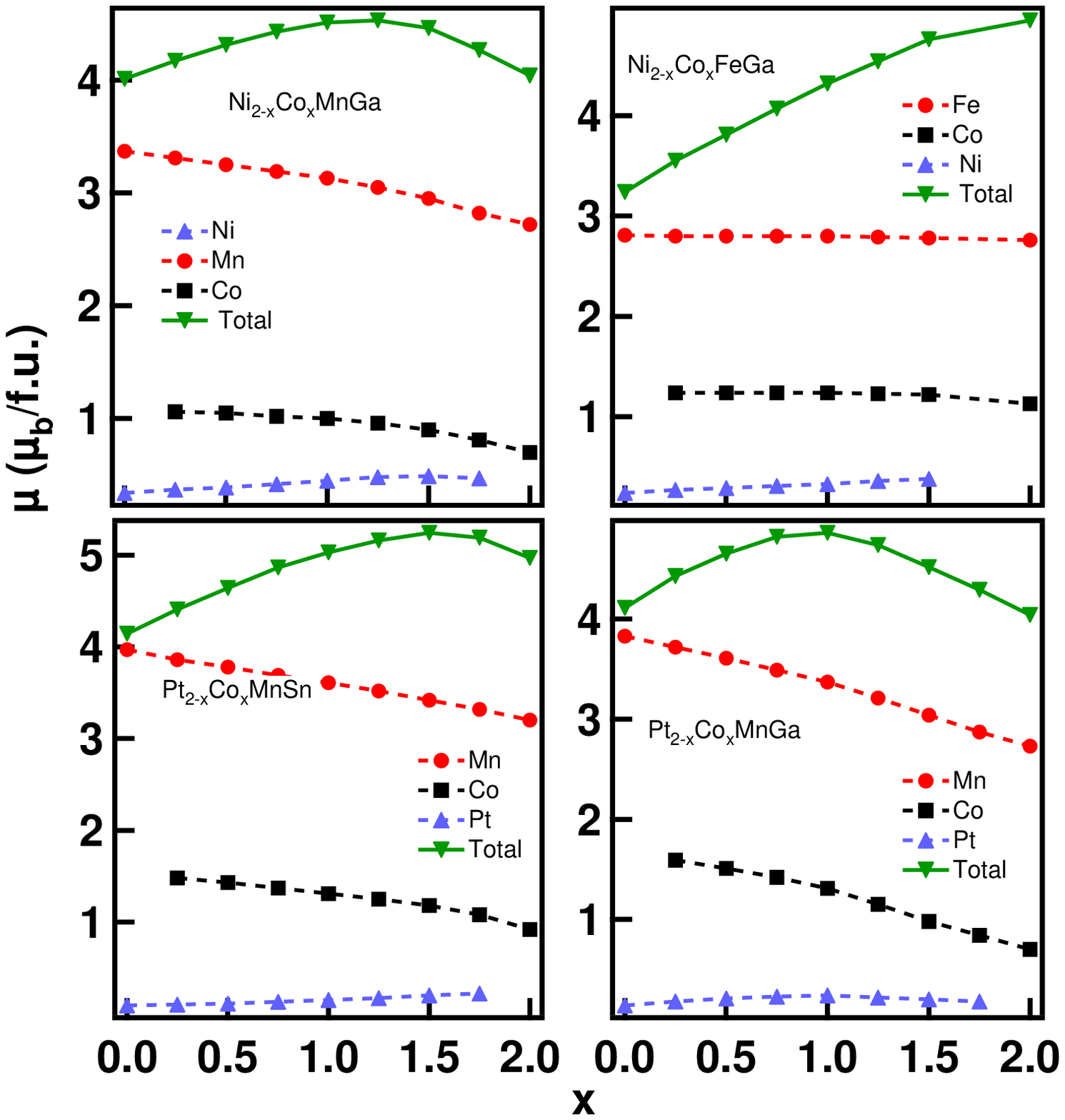}
\caption
{x dependence of magnetic moments for Ni$_{2-x}$Co$_{x}$MnGa, Ni$_{2-x}$Co$_{x}$FeGa, Pt$_{2-x}$Co$_{x}$MnSn, Pt$_{2-x}$Co$_{x}$MnGa.} 
\label{fig:1}
\end{figure}
\begin{figure}[ht*]
\includegraphics[width=8cm, height=8cm]{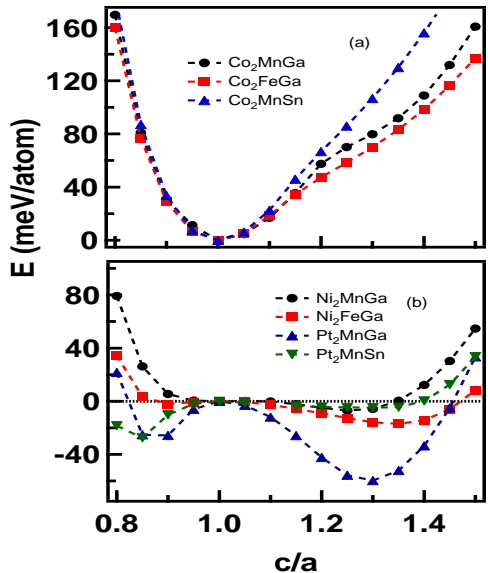}
\caption
{Variation of the total energy of  (a) Co$_{2}$MnGa,  Co$_{2}$FeGa, Co$_{2}$MnSn (b) Ni$_{2}$MnGa, Ni$_{2}$FeGa, Pt$_{2}$MnGa, Pt$_{2}$MnSn in their respective ground state magnetic configurations as a function of $c$/$a$.  Energy $E$ in the Y-axis signifies the energy difference between the cubic and tetragonal phase. Some results of these figures are part of published literature.\cite{prb-troy-2016,jmmm-troy-2016}}
\label{fig:1}
\end{figure}
\section{Results and Discussion} 

\textbf{\textit{Total and partial moments}}
 As mentioned above, we studied here four sets of materials, Ni$_{2-x}$Co$_x$MnGa, Ni$_{2-x}$Co$_x$FeGa, Pt$_{2-x}$Co$_x$MnGa, Pt$_{2-x}$Co$_x$MnSn with x=0.00, 0.25, 0.50, 0.75, 1.25, 1.50, 1.75, 2.00. At the two ends of the composition, i.e. for x=0.00 and x=2.00, all the materials except Pt$_{2}$MnSn already exist in the literature. In this study, all the materials corresponding to x=0.00 are likely to exhibit martensite transition. We predict here that Pt$_{2}$MnSn also possesses conventional Heusler alloy structure in its ground state and exhibits the martensite transition. All the studied materials here are ferromagnetic in nature. 

We observe from Figure-1 that for the three sets of materials namely, Ni$_{2-x}$Co$_{x}$MnGa, Pt$_{2-x}$Co$_{x}$MnGa, Pt$_{2-x}$Co$_{x}$MnSn the variation of total moment($\mu_{T}$) follows the same trend, which is for lower value of x, $\mu_{T}$ increases and then starts to fall at a higher x value, attaining a maximum value in between the range of x=0.00 to x=2.00. For Ni$_{2-x}$Co$_x$MnGa, the nature of variation of the moment as a function of x matches with the existing literature.\cite{prb-kanomata-2009}

The variation of the total moment as a function of x, can be well understood from the variation of the partial moments for the respective systems. We find that for Ni$_{2-x}$Co$_{x}$MnGa, Pt$_{2-x}$Co$_{x}$MnGa, Pt$_{2-x}$Co$_{x}$MnSn, the partial moment of Co and Mn-atom decreases linearly as a function of x. This may be because, as we move towards the higher value of x, the lattice parameter of the systems decreases which leads to decrease of the Mn and Co partial moment. But as the absolute value of moment of Co-atom is much larger compared than that of Ni or Pt, the total moment increases initially with increasing value of x. However, this increasing factor has to compete with the continuous reduction of  the partial moments of Co and Mn-atom, which dominates at higher value of x. This results in a fall of the total value of the moment. Because of these two competing factors, initially we get a maximum value of  $\mu_{T}$ and then it falls, finally reaches a value, very close to an integer following the Slater Pauling rule.\cite{JPD40HCK}

However, the total magnetic moment of  Ni$_{2-x}$Co$_{x}$FeGa increases linearly as a function of x. This type of variation may be because of the almost constant partial moment of Fe and Co atom over the entire range of x. This is probably due to the fact that the lattice parameters for the two end materials Ni$_{2}$FeGa (a= 5.76 \AA) and Co$_{2}$FeGa (a=5.73 \AA) are very close. Here the only controlling factor is the change of moment due to Ni substitution by Co-atom, which is always positive and proportional to the substitution and effectively results in a linear increase of the total moments of this system.

\textbf{\textit{Energy vs $c/a$ curve}}
Heusler alloys may be used as shape memory device if they undergo a structural transition from high temperature cubic phase to low temperature non-cubic phase upon cooling. The alloys, which are likely to undergo this structural transition, they must have the non-cubic phase with much lower energy compared to its cubic phase. We have applied a tetragonal distortion on the cubic phase of the stoichiometric material to probe whether they are favourable to undergo tetragonal distortion or not. In the upper panel of the Figure-2, we find that there is no lowering of energy under tetragonal distortion. For this set of materials, namely Co$_{2}$MnGa, Co$_{2}$MnSn, Co$_{2}$FeGa, the cubic phase is the lowest energy state ($c/a=1$) and they are not likely to undergo martensite transition.\cite{jalcom-troy-2015,prb-troy-2016,thesis-Antje}

In the lower panel of the Figure-2 we observe that for all the materials shown here (i.e. Ni$_{2}$MnGa, Ni$_{2}$FeGa, Pt$_{2}$MnGa, Pt$_{2}$MnSn), energy of the systems is lowered under tetragonal distortion which indicates to a possibility of martensite transition for these materials. Except Pt$_{2}$MnSn, the other three materials, namely Ni$_{2}$MnGa, Ni$_{2}$FeGa, Pt$_{2}$MnGa, of the lower panel are already reported to undergo martensite transition.\cite{jpcm-brown-1999,apl-liu-2003,apl-mario-2011}

\begin{figure}[ht*]
\includegraphics[width=8cm, height=8cm]{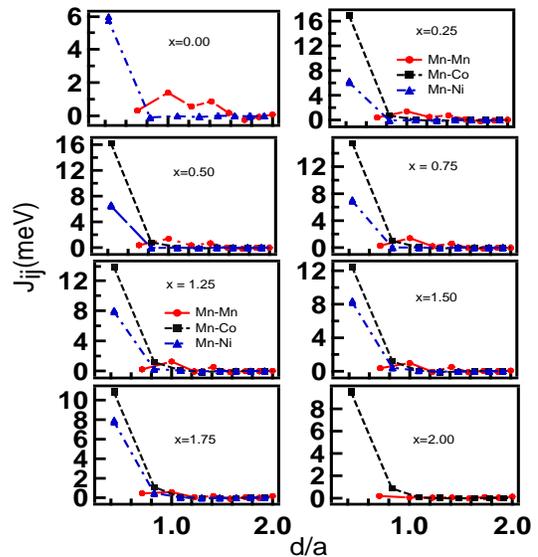}
\caption
{J$_{ij}$ of Mn atom with its neighbours as a function of normalized distance $d/a$ for Ni$_{2-x}$Co$_x$MnGa system. $a$ is the lattice parameter for x=0.00, 0.25, 0.50, 0.75, 1.25, 1.50, 1.75, 2.00.} 
\label{fig:1}
\end{figure}

\begin{figure}[ht*]
\includegraphics[width=8cm, height=8cm]{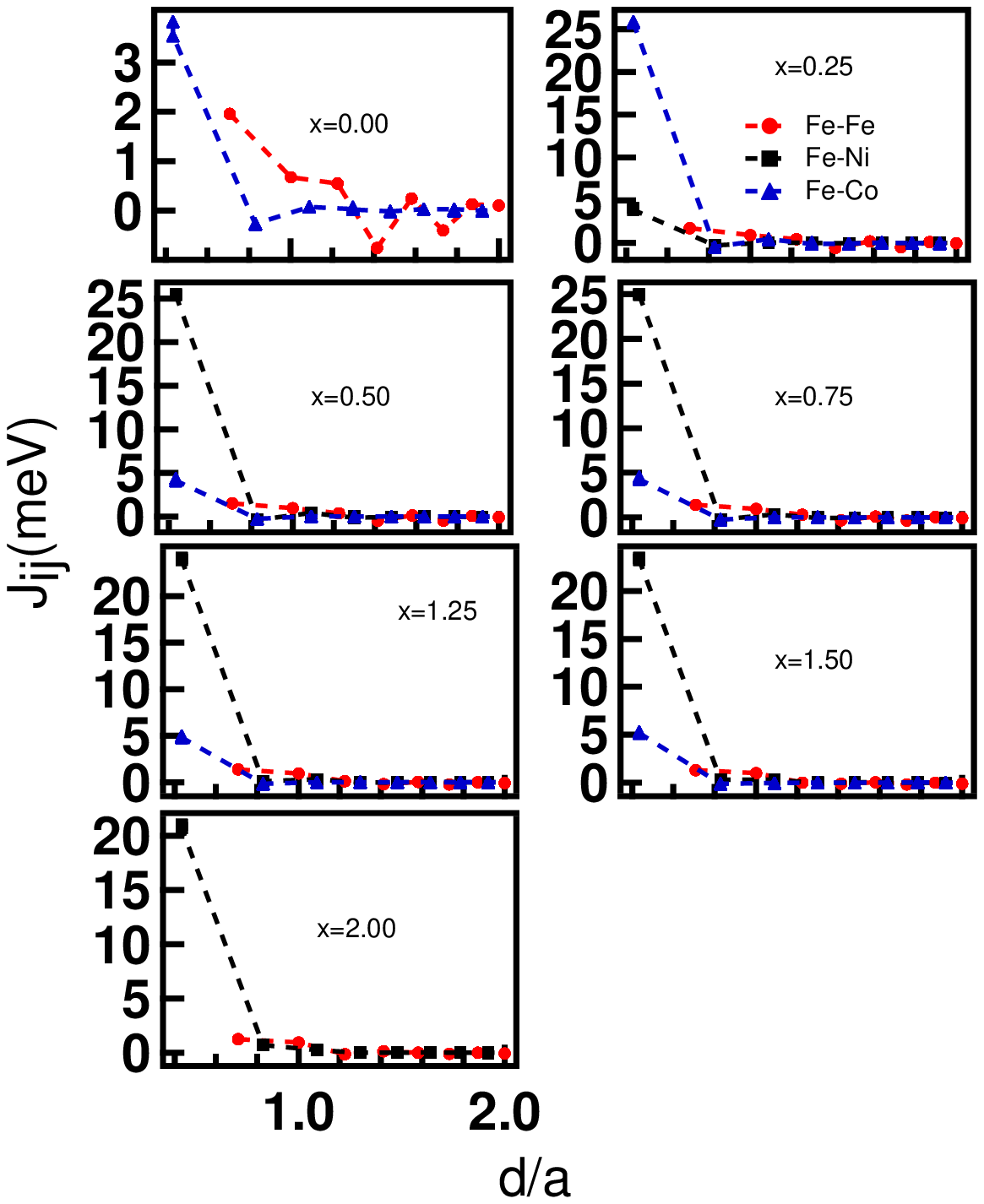}
\caption
{J$_{ij}$ of Fe atom with its neighbours as a function of normalized distance $d/a$ for Ni$_{2-x}$Co$_x$FeGa system. $a$ is the lattice parameter for x=0.00, 0.25, 0.50, 0.75, 1.25, 1.50, 2.00.} 
\label{fig:1}
\end{figure}

\begin{figure}[ht*]
\includegraphics[width=8cm, height=8cm]{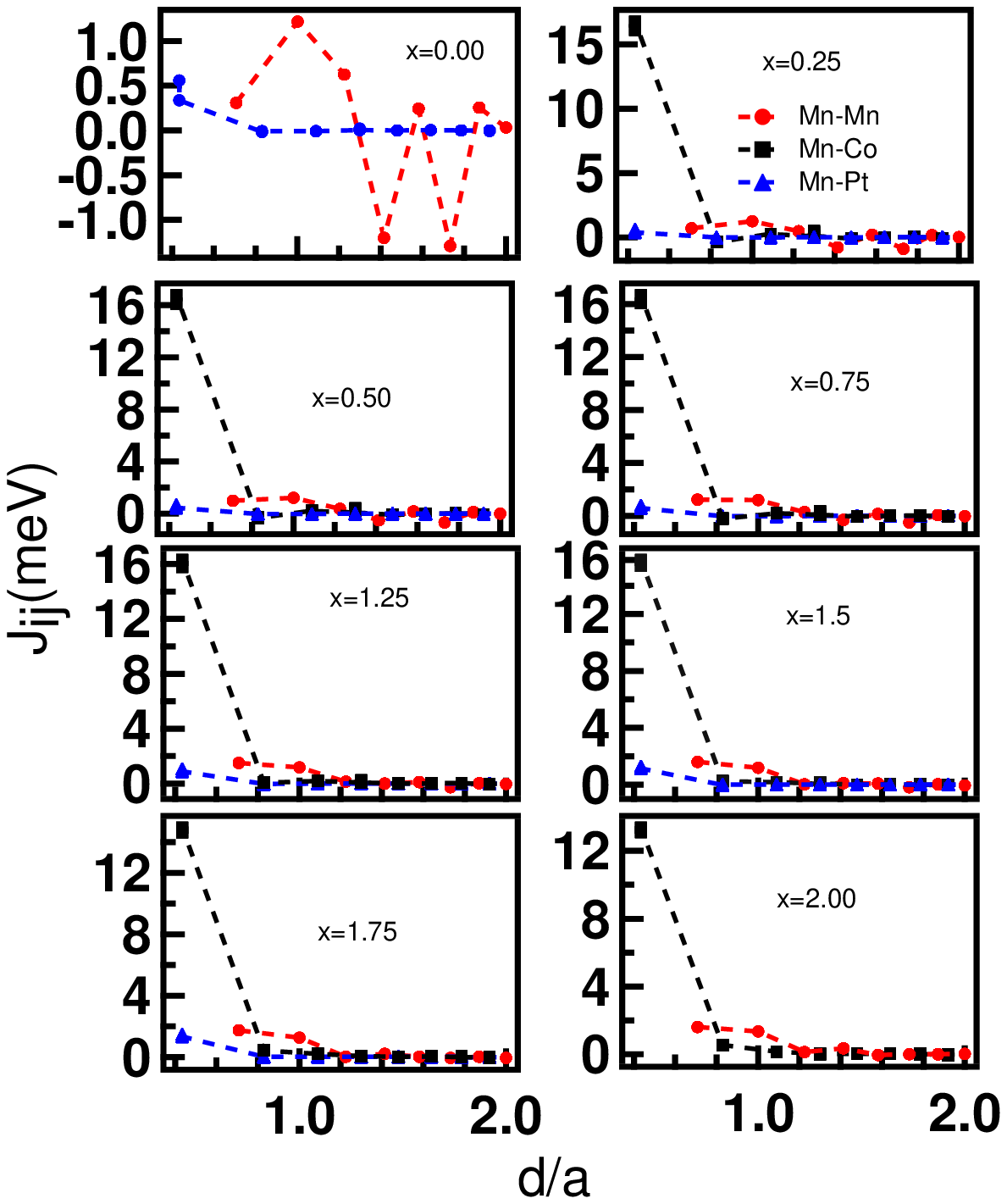}
\caption
{J$_{ij}$ of Mn atom with its neighbours as a function of normalized distance $d/a$ for Pt$_{2-x}$Co$_x$MnSn system. $a$ is the lattice parameter for x=0.00, 0.25, 0.50, 0.75, 1.25, 1.50, 1.75, 2.00.} 
\label{fig:1}
\end{figure}

\begin{figure}[ht*]
\includegraphics[width=8cm, height=8cm]{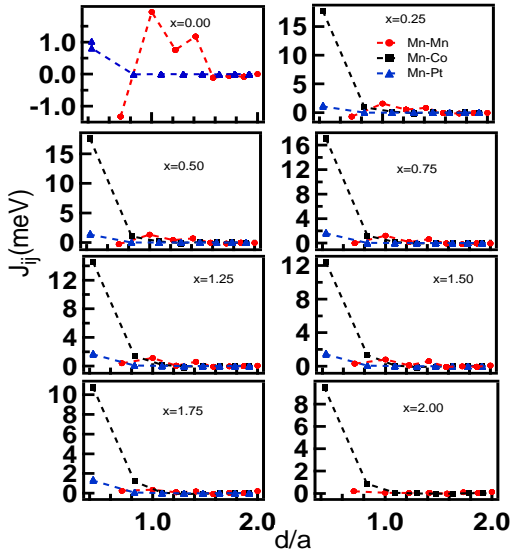}
\caption
{J$_{ij}$ of Mn atom with its neighbours as a function of normalized distance $d/a$ for Pt$_{2-x}$Co$_x$MnGa system. $a$ is the lattice parameter for x=0.00, 0.25, 0.50, 0.75, 1.25, 1.50, 1.75, 2.00.} 
\label{fig:1}
\end{figure}

Now we plot (Figure-3 to Figure-6) the Heisenberg exchange coupling parameters (J$_{ij}$), between Mn or Fe with other magnetic atoms of Ni$_{2-x}$Co$_x$MnGa, Ni$_{2-x}$Co$_x$FeGa, Pt$_{2-x}$Co$_x$MnGa, Pt$_{2-x}$Co$_x$MnSn (x=0.00, 0.25, 0.50, 0.75, 1.25, 1.50, 1.75, 2.00), as a function of interatomic spacing in the units of lattice parameter(a). We study the evolution of the magnetic interaction in going from a material which shows the SMA property (x=0.00) to  one which does not show the SMA property (x=2.00).

From Figure-3, we observe that the strength of the direct exchange interaction between Mn and Co is maximum  at x=0.25, i.e. for Ni$_{1.75}$Co$_{0.25}$MnGa and it is minimum for x=2.00 i.e. for Co$_{2}$MnGa. We observe that for the value of x=0.25, the partial magnetic moments of  Mn and Co atom  attain their maximum value among all the intermediate compounds (x=0.00 cannot be considered because there is no Co atom) which leads to the strongest direct exchange interaction between Mn and Co atom. The  scenario is exactly opposite for x=2.00.

For the other three sets of materials also, i.e. Ni$_{2-x}$Co$_x$FeGa, Pt$_{2-x}$Co$_x$MnGa, Pt$_{2-x}$Co$_x$MnSn, we find similar types variation of direct exchange interaction between Co and the $B$ atom ($B$=Mn for Pt$_{2-x}$Co$_x$MnGa, Pt$_{2-x}$Co$_x$MnSn and Fe for Ni$_{2-x}$Co$_x$FeGa).
which has been shown in Figure-4 to Figure-6. For all the materials studied here, we find that the exchange interaction energy between the $B$ and Co atom is much more stronger compared to $B$ and Ni (for Ni$_{2-x}$Co$_x$MnGa, Ni$_{2-x}$Co$_x$FeGa) or Pt atom (for Pt$_{2-x}$Co$_x$MnGa, Pt$_{2-x}$Co$_x$MnSn). This is because the partial moment of Co-atom is much higher compared to that of Ni or Pt-atom.

\begin{figure}[ht*]
\includegraphics[width=8cm, height=8cm]{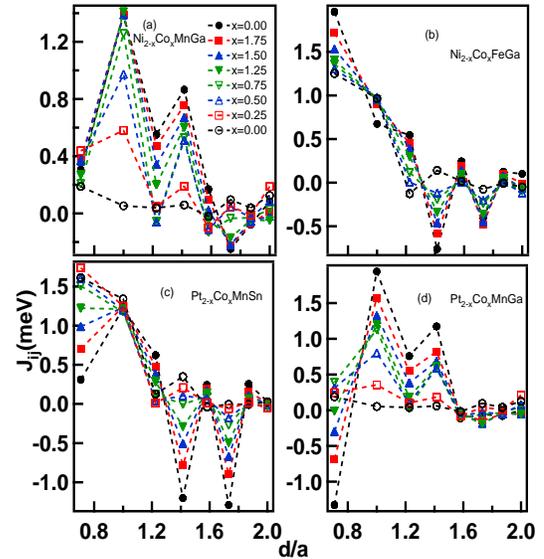}
\caption
{J$_{B-B}$ (B=Mn or Fe depending on the systems)  as a function of normalized distance $d/a$ for (a)Ni$_{2-x}$Co$_x$MnGa (b) Ni$_{2-x}$Co$_x$FeGa (c)Pt$_{2-x}$Co$_x$MnSn (d)Pt$_{2-x}$Co$_x$MnGa . $a$ is the lattice parameter for different values of x.} 
\label{fig:1}
\end{figure}

RKKY type of interaction plays a very important role in the systems where the localized moments are far apart to have any direct exchange interaction. There are an extensive studies on the RKKY interactions in various dilute magnetic systems, where the  magnetic atoms like Mn or Fe are present in a very low concentration in the nonmagnetic metallic host material.\cite{prb-smith-1976} The presence of RKKY interaction between  localized-like moments (Mn or Fe) was reported. This interaction was via the conduction electrons of the host material, which may be Au, Ag, Mo, Zn etc. Not only in metallic system, RKKY interaction plays a crucial role in determining the magnetic property of dilute magnetic semiconductor also.\cite{Priour-prl-2004} In this study all the systems contain Mn or Fe atom, and the magnetic moments are mainly confined to them. As all the systems studied here possess conventional Heusler alloy structure (A$_{2}$BC), the separation between the B atoms (Mn or Fe depending on the systems) are large enough to have a direct exchange interaction between them. For this kind of Heusler alloy structure B atom is surrounded by eight A atoms, which makes A atoms to play a very important role in determining the magnetic exchange interactions between B atoms themselves. The strong local nature of the magnetic moment of the B atoms spin-polarizes the free like electrons present in the system and the spin-polarized conduction electrons effectively couple the B atom.\cite{stearns-jap-1979} Previously in literature \cite{PRB28JK} it was mentioned for A$_{2}$MnC systems (A=Cu, Pd; C=Al, In, Sb), that the conduction electrons of the C atom take role in the coupling between Mn atoms. But in a recent study\cite{prb-sasioglu-2008} the role of conduction electron of A atom has also been confirmed for a number of Mn-based Heusler alloys. In our studied systems here, for a given series of materials,  C-atom is fixed which is Ga for 
Ni$_{2-x}$Co$_x$MnGa, Ni$_{2-x}$Co$_x$FeGa, Pt$_{2-x}$Co$_x$MnGa and Sn for Pt$_{2-x}$Co$_x$MnSn. However with substitution, nature of A atom changes. Here we will discuss only about the role of A-atom in the RKKY interaction between B atoms themselves. It is to be noted that the spin polarization of the conduction electron will depend on the local magnetic moment of the B atom and the number of the conduction electrons present in the system.
Now as we move from Ni$_{2}$MnGa to Co$_{2}$MnGa we are effectively reducing the number of conduction electrons of the system, as Ni has one more d-electron compared to Co-atom. This may cause a weaker coupling between the Mn atoms themselves. From Figure-7(a) we find that for Ni$_{2}$MnGa (x=0.00) the Mn-Mn interaction is the most oscillatory in nature (Heisenberg exchange coupling constant varies between 1.39 eV to -0.25 eV at $d/a=1$ and $d/a=1.73$) whereas for Co$_{2}$MnGa the oscillation is minimal (varies between 0.2 eV to -0.02 eV at $d/a=0.71$ and $d/a=1.58$). The strength of the oscillation reduces gradually as we move from Ni$_{2}$MnGa to Co$_{2}$MnGa.

For Ni$_{2-x}$Co$_{x}$FeGa system also we observe same kind of variation for Mn-Mn interaction as we move from x=0.00 to x=2.00. For x=0.00 i.e. for Ni$_{2}$FeGa the amplitude of Fe-Fe RKKY interaction varies between 1.96 eV ($d/a=0.71$) and -0.76 eV ($d/a=1.41$) which is the strongest among the Ni$_{2-x}$Co$_{x}$FeGa series.

In going from Pt-based systems to Co-based systems (Figure-7(c) and Figure-7(d)) also,  we are reducing the number of  conduction electrons.  One more factor which we must consider when we discus about Pt$_{2-x}$Co$_{x}$MnSn and Pt$_{2-x}$Co$_{x}$MnGa, is the change in lattice parameter between the compounds corresponding to x=0.00 and 2.00. On the other hand, for Ni$_{2-x}$Co$_{x}$MnGa  and Ni$_{2-x}$Co$_{x}$FeGa this change is very nominal as both Ni and Co has very close atomic radius. But as we move from Pt$_{2}$MnSn (a=6.46 \AA) to Co$_{2}$MnSn (5.98 \AA) there is a contraction of lattice parameter of about 0.48 \AA. For Pt$_{2}$MnGa (a=6.23 \AA)  to Co$_{2}$MnGa (a=5.72 \AA), a contraction of about 0.51 \AA takes place. This larger lattice parameter for Pt-based systems causes more localization of Mn partial magnetic moment (3.97 $\mu_{B}$ and 3.82 $\mu_{B}$ in Pt$_{2}$MnSn and Pt$_{2}$MnGa respectively) compared to the values in Co-based system (3.19 $\mu_{B}$ and 2.73 $\mu_{B}$ in case of Co$_{2}$MnSn and Co$_{2}$MnGa respectively). In Ref\cite{prb-bose-2011} Bose etal have mentioned that the strength of exchange interaction between two interacting magnetic moments also depends on value of the respective magnetic moments. 
Therefore, if we focus on Figure-7(c) we observe that the Mn-Mn exchange interaction energy  for x=0.00 (Pt$_{2}$MnSn) oscillates between a maximum value of 1.21 eV ($d/a=1.00$) and minimum value of -1.29 eV ($d/a=1.73$) but oscillation becomes weaker gradually as we increase x and for Co$_{2}$MnSn it varies between 1.61 eV ($d/a=0.71$) and 0.03 eV ($d/a=0.58$). It means RKKY type of interaction is much more strong in Pt$_{2}$MnSn compared to Co$_{2}$MnSn, which may be because of more localized-like Mn-moments in  Pt$_{2}$MnSn.

For Pt$_{2-x}$Co$_{x}$MnGa system also we find that the for x=0.00, RKKY type of interaction between Mn-Mn is the most oscillatory (for Pt$_{2}$MnGa it varies between -1.33 eV and  1.94 eV at $d/a =0.71, 1.00$ respectively) and gradually with increasing x, the interaction becomes less oscillating nature.

\section{Conclusion}

From density functional theory based calculations we study the effects of Co substitution in Ni and Pt-based Heusler alloys which are likely to show SMA.  Our results suggest that there is a decrease in strength of the RKKY interaction as we increase the Co doping at Ni or Pt site. It indicates about the dominant role played by A atom's d-electron in the formation of coupling between localized moments of B atom in the A$_{2}$BC system studied here. We also report the strong dependence of the strength of the RKKY interaction on the localization of B atom's magnetic moment. Our study signifies the implicit and important presence of  RKKY interaction  in the magnetic shape memory Heusler alloys.
\section{Acknowledgement}
Authors thank P. A. Naik and A. Banerjee for encouragement throughout 
the work. Authors 
thank S. R. Barman and C. Kamal for useful discussion and Computer 
Centre, RRCAT for technical support. TR thanks HBNI, RRCAT for 
financial support.


\begin{thebibliography}{}
\bibitem{phil-web-1984} P. J. Webster, K. R. A. Ziebeck, S. L. Town, M. S. Peak, Phil. Mag. B, {\bf 49}, 295 (1984).
\bibitem{apl-ullakko-1996} K. Ullakko, J. K. Huang, C. Kantner, R. C. O’Handley, and
V. V. Kokorin ,Appl. Phys. Lett. {\bf 69},1966(1996).
\bibitem{apl-sozinov-2002} A. Sozinov, A. A. Likhachev, N. Lanska, K. Ullakko, Appl Phys. Lett. {\bf 80}, 1746 (2002).
\bibitem{prb-sroy-2009} S. Roy, E. Blackburn, S. M. Valvidares, M. R. Fitzsimmons, S. C. Vogel, M. Khan, I. Dubenko, S. Stadler, N. Ali, S. K. Sinha, J. B. Kortright, Phys.Rev.B \textbf{79}, 235127 (2009).
\bibitem{prb-kataoka-2010} M. Kataoka, K. Endo, N. Kudo, T. Kanomata, H. Nishihara,
T. Shishido, R. Y. Umetsu, M. Nagasako, and R. Kainuma,Phys.
Rev. B {\bf 82}, 214423 (2010).
\bibitem{prb-achakrabarti-2005} A. Chakrabarti, C. Biswas, S. Banik, R. S. Dhaka, A. K. Shukla, S. R. Barman,Phys. Rev. B {\bf 72}, 073103 (2005).
\bibitem{prb-barman-2008} S. R. Barman, A. Chakrabarti, S. Singh, S. Banik, S. Bhardwaj, P. L. Paulose, B. A. Chalke, A. K. Panda, A. Mitra, A. M. Awasthi, Phys.Rev.B {\bf 78}, 134406 (2008).
\bibitem{apl-achakrabarti-2009} A. Chakrabarti, S. R. Barman, Appl. Phys. Lett. {\bf 94}, 161908 (2009)
\bibitem{jpcm-khan-2004} M. Khan, I. Dubenko, S. Stadler, and N. Ali,J. Phys.: Condens.  Matter {\bf 16}, 5259 (2004).
\bibitem{apl-mario-2011} M. Siewert, M. E. Gruner, A. Dannenberg, A. Chakrabarti, H. C.  Herper, M. Wuttig, S. R. Barman, S. Singh, A. Al-Zubi, T. Hickel, J. Neugebauer, M. Gillessen, R. Dronskowski, P. Entel, Appl.  Phys. Lett. {\bf 99}, 191904 (2011).
\bibitem{apl-stadler-2006} S. Stadler, M. Khan, J. Mitchell, N. Ali, A. M. Gomes, I. Dubenko, A. Y. Takeuchi, A. P. Guimaraes, Appl. Phys. Lett. {\bf 88}, 192511 (2006).
\bibitem{prb-achakrabarti-2013}A. Chakrabarti, M. Siewert, T. Roy, K. Mondal, A. Banerjee, M. E. Gruner, P. Entel, Phys. Rev. B {\bf 88},  174116 (2013) and references therein.
\bibitem{jalcom-troy-2015} T. Roy, M. E. Gruner, P. Entel, A. Chakrabarti, J. Alloys Compd. {\bf 632},  822 (2015).
\bibitem{prb-troy-2016} T. Roy, D. Pandey, A. Chakrabarti, Phys. Rev. B {\bf 93}, 184102, (2016).
\bibitem{jmmm-troy-2016} T. Roy, A. Chakrabarti, J. Magn. Magn. Mater. \textbf{401}, 929 (2016).
\bibitem{jmmm-troy-2017} T. Roy, A. Chakrabarti, J. Magn. Magn. Mater. {\bf 423}, 395, (2017).
\bibitem{prl-Groot-1983}R. A. de Groot, F. M. Mueller, P. G. van Engen, K. H. J. Buschow, Phys. Rev. Lett., {\bf 50}, 2024 (1983).
\bibitem{JPD40HCK}  H. C. Kandpal, G. H. Fecher, C. Felser, J. Phys. D: Appl. Phys., {\bf 40}, 1507 (2007).


\bibitem{PRB-76-024414-2007} J. K\"ubler, G. H. Fecher, C. Felser, Phys. Rev. B \textbf{76}, 024414 (2007).
\bibitem{PRB28JK} J. K\"ubler, A. R. Williams, C. B. Sommers, Phys. Rev. B, {\bf 28}, 1745 (1983).
\bibitem{RKKY}M. A. Ruderman and C. Kittel, Phys. Rev., {\bf 96}, 99 (1954); T. Kasuya, Prog. Theor. Phys., {\bf 16}, 45 (1956); K. Yosida, Phys. Rev., {\bf 106}, 893 (1957).  
\bibitem{prb-sasioglu-2008} E. \ifmmode \mbox{\c{S}}\else \c{S}\fi{}a\ifmmode \mbox{\c{s}}\else \c{s}\fi{}\ifmmode \imath \else \i \fi{}o\ifmmode \breve{g}\else \u{g}\fi{}lu, L. M. Sandratskii, P. Bruno, Phys Rev B {\bf 77}, 064417, (2008).
\bibitem{staunton-jpcssp-1988} J. B. Staunton, B. L. Gyorffy, J. Poulter, P. Strange, J. Phys. C: Solid State Phys. {\bf 21}, 1595, (1988).
\bibitem{prb-kreese-1996}G. Kresse, J. Furthm\"uller, Phys. Rev. B, {\bf 54}, 11169 (1996); G. Kresse, D. Joubert, ~Phys. Rev. B, {\bf 59}, 1758 (1999); VASP 5.2 programme package is fully integrated in the MedeA® platform (Materials Design, Inc.) with a graphical user interface enabling the computation of the properties.
\bibitem{prb-blochl-1994} P. E. Blochl, Phys. Rev. B \textbf{50}, 17953 (1994).
\bibitem{rep-ebert-2011} H. Ebert, D. Kodderitzsch, J. Minar, Rep. Prog. Phys. \textbf{74}, 096501 (2011).
\bibitem{LDA} S.H. Vosko, L. Wilk, and M. Nusair, Can. J. Phys.{\bf 58}, 1200
(1980).
\bibitem{prb-kanomata-2009} T. Kanomata, Y. Kitsunai, K. Sano, Y. Furutani, H. Nishihara, R. Y. Umetsu, R. Kainuma, Y. Miura, M. Shirai, Phys. Rev. B,  {\bf 80}, 214402, (2009).

\bibitem{thesis-Antje} Antje Dannenberg, Ab initio and Monte Carlo investigations of structural,
electronic and magnetic properties of new ferromagnetic Heusler alloys with
high Curie temperatures (Ph.D. thesis), University of Duisburg-Essen, 2011.
\bibitem{jpcm-brown-1999}P.J. Brown, A.Y. Bargawi, J. Crangle, K.-U. Neumann, K. Ziebeck, J. Phys.:
Condens. Matter {\bf 11} 4715  (1999).
\bibitem{apl-liu-2003} Z.H. Liu, M. Zhang, Y.T. Cui, Y.Q. Zhou, W.H. Wang, G.H. Wu, Appl. Phys. Lett. {\bf 82}
 424 (2003).
\bibitem{prb-smith-1976} F. W. Smith, Phys. Rev. B, {\bf 14}, 241, (1976). F. W. Smith, Phys. Rev. B, {\bf 13}, 2976, (1976). F. W. Smith, Phys. Rev. B, {\bf 10}, 2034, (1974).  F. W. Smith and M. P. Sarachik,  Phys. Rev. B, {\bf 16}, 4142, 1977.
\bibitem{Priour-prl-2004} D. J. Priour, Jr., E. H. Hwang, S. Das Sarma, Phys. Rev. Lett. {\bf 92}, 117201, (2004).
\bibitem{stearns-jap-1979}M. B. Stearns, J. Appl. Phys.{\bf 50}, 2060 (1979).
\bibitem{prb-bose-2011} S. K. Bose, J. Kudrnovsk'y, V. Drchal, I. Turek, Phys. Rev. B, {\bf 84}, 174422 (2011).



\end{thebibliography}
\end{document}